# Plethora of tunable Weyl fermions in kagome magnet $Fe_3Sn_2$ thin films

Zheng Ren[1]*, Hong Li[1]*, Shrinkhala Sharma[1]*, Dipak Bhattarai[2], He Zhao[1], Bryan Rachmilowitz[1], Faranak Bahrami[1], Fazel Tafti[1], Shiang Fang[3], Madhav Ghimire[2], Ziqiang Wang[1] and Ilija Zeljkovic[1]

[1] Department of Physics, Boston College, Chestnut Hill, MA 02467
[2] Central Department of Physics, Tribhuvan University, Kirtipur, 44613, Kathmandu, Nepal
[3] Department of Physics, Massachusetts Institute of Technology, Cambridge, MA 02139, USA
* equal contribution

**Abstract**

**Interplay of magnetism and electronic band topology in unconventional magnets enables the creation and fine control of novel electronic phenomena. In this work, we use scanning tunneling microscopy and spectroscopy to study thin films of a prototypical kagome magnet $Fe_3Sn_2$. Our experiments reveal an unusually large number of densely-spaced spectroscopic features straddling the Fermi level. These are consistent with signatures of low-energy Weyl fermions and associated topological Fermi arc surface states predicted by theory. By measuring their response as a function of magnetic field, we discover a pronounced evolution in energy tied to the magnetization direction. Electron scattering and interference imaging further demonstrates the tunable nature of a subset of related electronic states. Our experiments provide the first visualization of how *in-situ* spin reorientation drives changes in the electronic density of states of the Weyl fermion band structure. Combined with previous reports of massive Dirac fermions, flat bands and electronic nematicity, our work establishes $Fe_3Sn_2$ as a unique platform that harbors an extraordinarily wide array of topological and correlated electron phenomena.**

Introduction

Weyl fermions are condensed matter analogues of the elusive Weyl particles in high energy physics [1–5]. They are characterized by distinct reciprocal space points where bulk electronic bands touch [6,7]. Emerging in pairs of two, Weyl crossings, or nodes, produce topological Fermi arc states at the surface that connect two Weyl points of opposite chirality [6]. A Weyl semimetal phase in a material can be realized by breaking either the inversion symmetry of the crystal or the time-reversal symmetry [6]. The Weyl phase rooted in the former has been experimentally realized in a series of transition metal monopnictides [8–10] and transition metal dichalcogenides [11–13]; the latter however, has only been observed more recently in a kagome ferromagnet $Co_3Sn_2S_2$ [14–17] and a few other select magnets [18–20].

Due to the intimate connection between magnetism and electronic structure, Weyl nodes in magnetic Weyl semimetals can in principle be manipulated by changing the underlying spin texture. In contrast to massive Dirac fermions in Chern magnets where spin reorientation modulates the band gaps [21,22], Weyl fermions in three dimensions in this scenario should in principle remain gapless. Interestingly however, spin reorientation can change the energy and momentum space position of Weyl nodes, modify the shape of the Fermi arcs and even create and annihilate pairs of Weyl points under some conditions [23,24]. Unconventional evolution of magnetoresistance [25] and anomalous Hall [25,26] measurements hinted at the tunable nature of the Weyl node structure. However, Weyl fermions observed in topological magnets thus far are typically sparse in energy, reside away from the Fermi level, and as such, a direct evidence for their controllable manipulation remains limited. Here, we use low-temperature scanning tunneling microscopy/spectroscopy (STM/S) to study kagome ferromagnet $Fe_3Sn_2$ thin films synthesized by molecular beam epitaxy (MBE). STM differential conductance (d$I$/d$V$) spectra reveal a series of densely-spaced spectral features in close proximity to the Fermi level, consistent with the abundance of Weyl fermions predicted in $Fe_3Sn_2$ [27]. By rotating the magnetization direction using an external magnetic field, we find that both d$I$/d$V$ spectra and d$I$/d$V$ maps show a noticeable change. This spectral evolution terminates as the magnetization direction is saturated along the crystal $c$-axis, demonstrating the magnetization-direction-driven nature of the observed phenomena. Our experiments provide strong evidence for the existence of large number of Weyl fermion clusters in the close vicinity of the Fermi level in $Fe_3Sn_2$, and reveal the elusive spectroscopic signature of Weyl fermions tunable by spin reorientation.

**Results**

Kagome materials are composed of two-dimensional layers characterized by the corner-sharing hexagonal lattice (kagome lattice). In the presence of spin-orbit coupling, electronic correlations and magnetism, layered kagome materials can host various novel electronic states, such as topological flat bands [28–31], Chern magnet phase [22], and Weyl nodes accompanied by Fermi arcs [14–16]. $Fe_3Sn_2$ is a prototypical kagome system [21,27,32,33] that has a hexagonal lattice structure ($a$=$b$=5.34 Å, $c$=19.79 Å) composed of alternating kagome $Fe_3Sn$ bilayers and hexagonal Sn layers (Fig. 1a). We study $Fe_3Sn_2$ thin films grown on $SrTiO_3$(111) substrates using molecular beam epitaxy (MBE) (Methods). The sharp streaky reflection high energy electron diffraction (RHEED) images of our films demonstrate a high quality of the surface (Fig. 1b). The crystal structure and the orientation of the films is confirmed by X-ray diffraction measurements, which show a $c$-axis diffraction peak at 41.1 degrees in the vicinity of, but distinctly separated from, the substrate peak at 40.0 degrees (Fig. 1c, Supplementary Note 1). Magnetization measurements further reveal the expected ferromagnetic behavior, with the saturation field of about 1 T and the saturation moment of 1.8 $\mu_B$ per Fe atom (Fig. 1f). This is in agreement with the equivalent properties reported in bulk single crystals of $Fe_3Sn_2$ that exhibit intrinsic in-plane magnetization [21,34]. As an out-of-plane magnetic field is applied, the magnetization can be fully polarized along the $c$-axis, exhibiting a similar saturation field and moment (Fig. 1f).

STM topographs display a hexagonal lattice with $a \approx b \approx 5.35$ Å lattice constants, consistent with the expected lattice structure of $Fe_3Sn_2$ (Fig. 1d). We focus on the $Fe_3Sn$ kagome surface termination shown in Fig. 1d, identified to be the kagome surface layer for the following reasons. First, the crystal structure of $Fe_3Sn_2$ consist of two $Fe_3Sn$ kagome planes stacked on top of one another (labeled A and B in Fig. 1a), which are expected to be offset by a π phase shift along the two in-plane lattice directions. This expected stacking sequence is experimentally confirmed in our films by imaging the lattice structure on both sides of the occasionally encountered step edge; while the lattice morphology in STM topographs appears nearly identical on the two terraces, there exists a half-unit-cell offset in the *ab*-plane between the two (Fig. 1e, Supplementary Note 2). Second, the morphology of the STM topograph also looks qualitatively similar to $Fe_3Sn$ kagome surface termination of bulk single crystals, and markedly different from the Sn surface [21,28] (Supplementary Note 3).

To gain insight into the electronic structure of the system, we turn to d*I*/d*V* spectroscopy. d*I*/d*V* spectra of our films (Fig. 2b) show a general consistency with d*I*/d*V* spectra of bulk single crystals [28]. In particular, we observe a density-of-states upturn starting at about -150 mV (Fig. 2b), which was previously attributed to a partial flat band around the same energy [28]. Fourier transforms (FTs) of normalized d*I*/d*V*(**r**,*V*) maps show diffuse scattering wave vectors primarily located near the center of the FT (inset in Fig. 2a). We plot the radially-averaged linecuts, starting at the FT center, as a function of energy (Fig. 2c). The most prominent feature observed is the enhancement in the scattering signal below $E_0 \approx$ -150 mV (Fig. 2c). Similar enhancement of the scattering signal below $E_0$ is observed in different films, with minimal changes to the energy of the feature (inset in Fig. 2c), thus demonstrating the robustness of the electronic structure of our films.

Next we focus on the spectroscopic properties near the Fermi level ($E_F$). Remarkably, d*I*/d*V* spectra exhibit a series of features in close proximity to $E_F$ (Fig. 3a). These are more clearly observed in numerical second derivatives of d*I*/d*V* spectra ($d^3I/dV^3$ curves) (Fig. 3b). Upon the application of magnetic field parallel to the *c*-axis, we find that these spectral features rapidly evolve (Fig. 3a-c). Their evolution plateaus at 1 T, and shows the same behavior when the field is reversed (applied antiparallel to *c*-axis) (Fig. 3e,f). As bulk magnetization, lying in-plane at zero field, gradually tilts and saturates in out-of-plane direction at 1 T (Fig. 1f), this points towards magnetization-direction-driven modification to the electronic density-of-states.

To quantify the evolution of spectral features, we determine the energy of each "dip" in $d^3I/dV^3$ curves (Fig. 3b), and plot it as a function of magnetic field (Fig. 3c). We note that a dip in a $d^3I/dV^3$ curve directly corresponds to a locally enhanced density-of-states in the corresponding d*I*/d*V* spectrum (Fig. 3a,b). The dispersion shows a complex field-induced behavior. First, we observe that some of the spectral features evolve to higher energy with field (i.e. $E_3$ and $E_6$ in Fig. 3c), while others gradually shift to lower energy ($E_1$, $E_2$, $E_4$ and $E_5$ in Fig. 3c). Second, the dispersion velocities are also highly variable; for example, $E_3$ changes its peak position by ~5.4 meV when the field is changed from 0 to 1 T, while the shift of $E_4$ is only ~2.3 meV in the same field range. Lastly, it is apparent that the relative sharpness of spectral peaks also changes with field. For instance, $E_3$ becomes significantly sharper approaching 1 T, while $E_2$ diffuses away.

Spectral peaks in STM dI/dV spectra of related materials have been typically associated with either Landau levels [35–37], flat bands [28,38], massive Dirac fermions [21] and/or saddle points [39]. We deem that these are unlikely to explain the emergence and the evolution of spectral features in our work for the following reasons. First, the absence of peak energy change above 1 T (Fig. 3a) rules out Landau levels as the root cause of our observations. We can also rule out that the spectral peaks in our data originate from Dirac mass acquisition and evolution [21] because: (1) the Dirac crossings at K points in $Fe_3Sn_2$ are located well below $E_f$, at -70 meV and -180 meV [32], and (2) bilayer splitting between Dirac cone energies is expected to be about 110 meV in ARPES [32] and 80 meV in optical measurements [40], which is significantly larger than the energy range where the dispersing features in our data are observed. The flat band identified in $Fe_3Sn_2$ is also about 150 meV below the Fermi level [28], consistent with its spectroscopic signature in our data (Fig. 2). Lastly, bulk electronic band structure of $Fe_3Sn_2$ is strongly dispersive along $k_z$ [41]. This in principle blurs out distinct bulk band features, such as for example saddle points, making them difficult to be observed by STM that averages over all $k_z$ momenta.

This brings up an intriguing question of the origin of the extraordinary number of spectral peaks located only several tens of millielectronVolts away from $E_f$. Assuming that the Fe moment is oriented along the *c*-axis, recent *ab initio* calculations suggest the existence of at least six sets of Weyl points within 50 meV of $E_F$ [27] (Fig. 3d). Upon closer inspection, the energies of calculated Weyl points are peculiarly consistent with the energies of the spectral features observed in our experiments (Fig. 3c,d). As the emergence of a pair of Weyl nodes is accompanied by the formation of a surface state Fermi arc connecting the two nodes, enhanced density-of-states in our d*I*/d*V* spectra can be understood as the formation of Fermi arcs that couple to the STM tip more significantly, leading to the enhanced density-of-states measured in STM d*I*/d*V* spectra. We note that Fermi arcs will also disperse with energy – the spectral peak in the momentum-integrated local DOS may cover all of the dispersion range, which may be relatively narrow given that some nodes here are expected to be Type-II (Supplementary Table 1), or a part of this range where the Fermi arc spectral weight is the most pronounced. In turn, the spectral peak position in dI/dV spectrum does not necessarily have to correspond exactly to the energy of the associated Weyl node. The enhancement of the local density-of-states in dI/dV spectra has previously been interpreted to arise due to Fermi arcs and associated Weyl nodes in for example magnetic Weyl semimetal $Co_3Sn_2S_2$ [42] or non-magnetic chiral crystal CoSi [43]. Similar to these, our data is consistent with the formation of Fermi arcs related to the underlying Weyl crossings. The fact that the d*I*/d*V* spectral evolution terminates once the magnetization is saturated (Fig. 3a-c) provides strong support in favor of a magnetization-direction-driven evolution. While increased quasiparticle lifetime closer to $E_F$ could in principle lead to spectral peak sharpening, we point out that $E_6$ actually gets more prominent as it moves away from $E_F$. Therefore, the evolution of spectral peak morphology cannot be explained by the change in quasiparticle lifetime by itself.

As momentum-space signature of a Fermi arc dispersion could in principle be detected in STM d*I*/d*V* maps [16,44], we proceed to look for such signatures in our data (Fig. 4). We observe a rapidly dispersing wave vector at $(Q_{Bragg,1}+Q_{Bragg,2})/3$ ($q_1$ denoted by green circle in Fig. 4a), which emerges just above $E_F$. The reciprocal-space extent of the wave vector becomes progressively larger with increased energy until it ultimately disappears at ~18 meV (Fig. 4h,i). Based on the hexagonal symmetry of the crystal structure, this is likely a reflection of scattering between electronic states at K and K' points. We mention that we also observe a scattering wave vector around the atomic Bragg peak in a comparable energy range, which could be consistent with scattering between states near two different K points (Supplementary Note 7 and Supplementary Figure 7). We note that in Ref. [27], one set of calculated Weyl points indeed resides at the Brillouin zone corners at 17 meV above $E_F$ (Fig. 4b,c). This is consistent with the energy range of the scattering wave vector $q_1$ (Fig. 4h,i). For pedagogical purposes, we connect the Weyl points of the opposite chirality with a simple arc (Fig. 4b,c) and examine the concomitant scattering signature (Fig. 4d-g). We hypothesize that the FT wave vector becomes broader concomitant with the change in the Fermi arc shape as a function of energy (Fig. 4b-g). We note that additional theoretical modeling will be necessary to determine the exact shape of the Fermi arcs and the one shown in Fig. 4b,c is a simple illustration that is consistent with our STM data. Similar to the field dispersion of d$I$/d$V$ spectra (Fig. 3), we find that the spectral profile of $q_1$ also changes with applied magnetic field (Fig. 4i). This directly indicates a change in the electronic state visualized in momentum space; transfer of the spectral weight can be more easily seen in the numerical difference between data at the two field magnitudes (Fig. 4i). We note that the dispersion velocity of $q_1$ observed in our data (Fig. 4i) is about 5 times smaller than that of Dirac fermions measured by ARPES [32] thus again pointing against Dirac dispersion being at the root of our observations.

**Discussion**

$Fe_3Sn_2$ is a rare topological magnetic system where numerous Weyl points have been predicted theoretically to reside in close proximity to $E_F$ [7,24,27]. Our experiments provide experimental visualization consistent with this unusual physical picture. We reveal a series of near-$E_F$ spectral features, which can be understood to arise from the enhanced density-of-states from topological Fermi arc surface states intimately related to the underlying Weyl points. By applying the magnetic field to tilt the magnetization direction, we detect their concomitant response as the spins are reoriented. Favorable positioning of these points in close proximity to $E_f$ could be exploited and further explored in high-resolution transport and opto-electronic measurements. As hypothesized in Ref. [41], a magnetic Weyl semimetal phase could be a plausible explanation for the observed anomalous Hall Effect in $Fe_3Sn_2$. It is also interesting to note that our measurements detect enhanced d$I$/d$V$ signal along the boundary of a kagome terrace, possibly indicating the existence of an edge mode (Supplementary Figure 4). Given the emergence of this mode below Fermi level in the vicinity of the expected Dirac node [32], this edge mode could be consistent with the chiral edge states expected to appear in the presence of massive Dirac fermions in $Fe_3Sn_2$. This provides another intriguing direction that can be pursued in future MBE-STM experiments of $Fe_3Sn_2$, aimed to study field tunability of edge modes in kagome magnets building on the foundation established by our experiments.

## Methods

**MBE growth.** Buffered hydrogen fluoride treated Nb-doped (0.05 wt%) $SrTiO_3$ (111) substrate (5 mm x 5 mm x 0.5 mm) (Shinkosha) was cleaned in acetone and 2-propanol in an ultrasonic bath and then introduced into our MBE system (Fermion Instruments) with a base pressure of ~ 5 x $10^{-10}$ Torr. The substrate was first slowly heated to the growth temperature at ~660°C, which was continuously monitored by a pyrometer (emissivity=0.7). Thereafter, Fe (99%) and Sn (99.9999%) were co-evaporated from individual Knudsen cells after the flux rates were calibrated using a quartz crystal microbalance (QCM). A flux ratio of Fe:Sn=1.5:1 was roughly achieved as the temperatures of Fe and Sn were set at 1187°C and 970°C, respectively. For STM measurements, thin films were quickly transferred using a vacuum suitcase chamber held at 5 x $10^{-11}$ Torr, and were never exposed to air. STM experiments shown in Figures 3 and 4 were performed on about 20 nm thick films. For *ex-situ* X-ray diffraction and magnetization measurements, a brief exposure to air and the storage in the desiccator over a course of few days did not seem to interfere with the sample properties.

**STM measurements.** STM data was acquired using a customized Unisoku USM1300 STM at the base temperature of ~4.5 K. Spectroscopic measurements were acquired using a standard lock-in technique at 915 Hz and bias excitations as detailed in the figure captions. STM tips were home-made chemically etched tungsten tips, annealed in UHV to bright orange color prior to STM imaging.

**X-ray diffraction.** X-ray diffraction measurements were carried out at room temperature in reflection (Bragg-Brentano) geometry using a Bruker D8 ECO diffractometer equipped with a copper $K_\alpha$ source, a nickel filter to absorb $K_\beta$ radiation, and a 2.5° Soller slits after the source and before the LYNXEYE XE 1D energy-dispersive detector. By comparing the $2\theta$ angles of the peaks with bulk $Fe_3Sn_2$, a *c*-axis strain less than 0.2% can be extracted.

**Magnetic measurements.** DC magnetization measurements were performed in a Quantum Design MPMS3. Samples were mounted on a small background quartz holder using rubber cement glue in order to obtain the $B_{||}$ measurements. In Fig. 1f, we first subtracted the substrate signal measured from an identical piece of bare substrate, then converted the raw magnetization measured by MPMS to the magnetic moment per Fe atom based on the volume of the film. The sample volume is approximately determined based on the time of growth and the expected elemental flux calibrated by QCM.

## Data availability

Raw STM data used for the analysis shown in Figs. 2-4 can be downloaded from: https://doi.org/10.5281/zenodo.6805185.

## Code availability

The computer code used for data analysis is available upon request from the corresponding author.

**Acknowledgements**

We are grateful to Joe Checkelsky for the insightful conversations and the use of computational resources. Z.W. acknowledges the support of U.S. Department of Energy, Basic Energy Sciences Grant No. DE-FG02-99ER45747 and the Cottrell SEED Award No. 27856 from Research Corporation for Science Advancement. S.F. acknowledges support by the Gordon and Betty Moore Foundation EPiQS Initiative, Grant No. GBMF9070 to Joe Checkelsky.

**Author contributions**

MBE growth was performed by Z.R and S.S. STM experiments were carried out by H.L and H.Z. Z.R. analyzed the STM data. Z.R. and B.R. performed magnetization measurements. X-ray measurements were done by F.B. and F.T. D.B, S.F. and M.G. performed theoretical calculations. Z.W. provided theoretical input on the interpretation of STM data. I.Z. supervised the project. I.Z, Z.R, Z.W, S.F. and M.G. wrote the manuscript with input from all the authors.

**Competing interests**

The authors declare no competing interests.

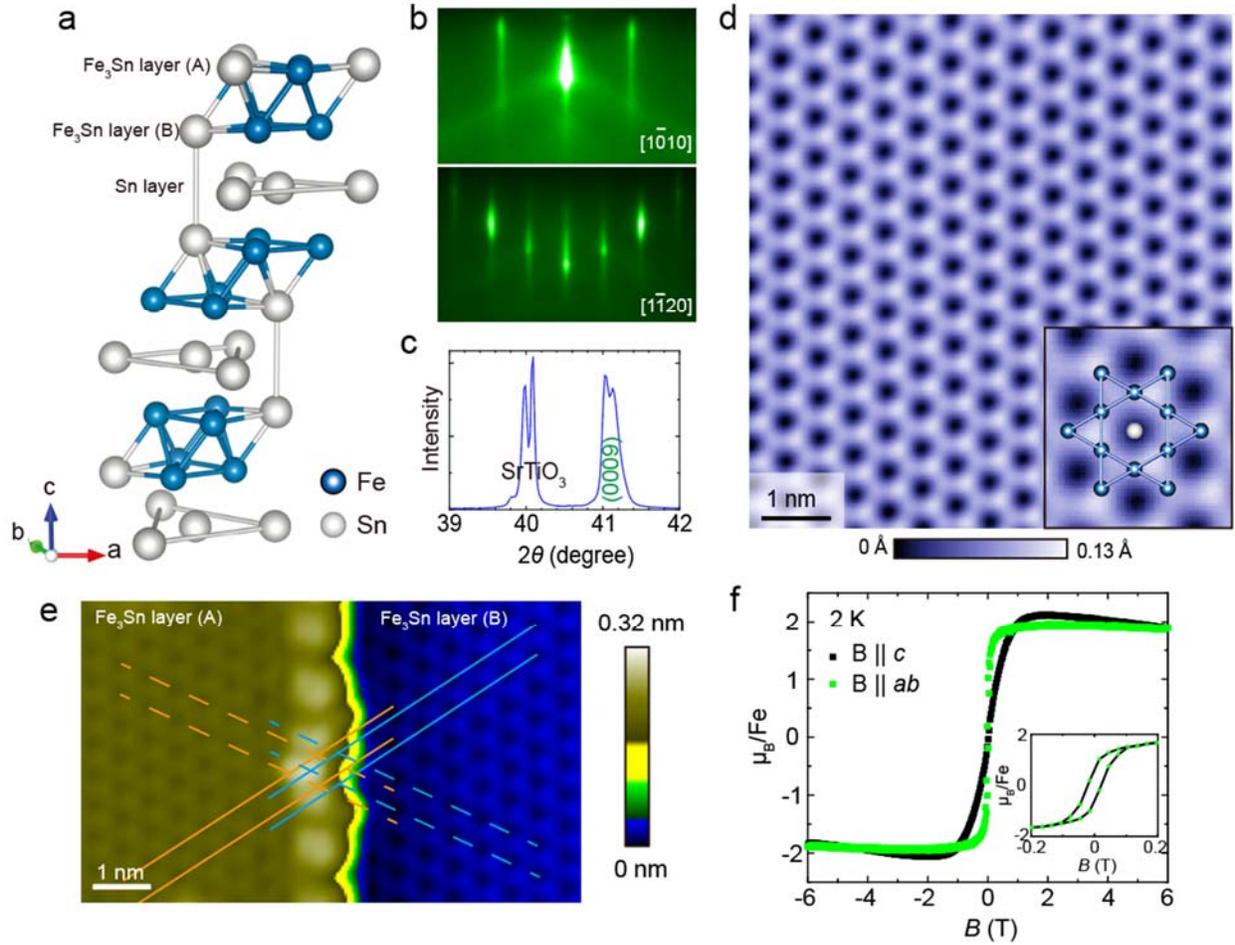

**Figure 1. Structural and magnetic characterization of Fe₃Sn₂ films. a** Crystal structure of one unit cell of Fe₃Sn₂. Blue and white spheres represent Fe and Sn atoms, respectively. **b** Post-growth RHEED images of Fe₃Sn₂ thin film along two high-symmetry in-plane directions. **c** XRD data of the sample showing the SrTiO₃ substrate peak and the Fe₃Sn₂ (0009) peak. **d** High-resolution STM topograph acquired at the Fe₃Sn surface. Inset in **d** magnifies a small region with the schematic of the kagome structure of Fe₃Sn layer superimposed. **e** STM topograph containing an Fe₃Sn-Fe₃Sn atomic step. Orange (blue) lines follow the nearest-neighbor Sn atoms in the upper (lower) Fe₃Sn layer (See Supplementary Note 2). Solid and dashed lines are parallel to the in-plane *a* and *b* lattice vectors, respectively. **f** Magnetization curves as a function of magnetic field applied in the *ab*-plane (green) and along the *c*-axis (black) after subtracting the magnetic moment from the substrate. Inset in **f** displays the hysteresis loop near 0 T at 2 K. STM setup conditions: **d** $I_{set}$ = 100 pA, $V_{sample}$ = 100 mV; **e** $I_{set}$ = 280 pA, $V_{sample}$ = 35 mV.

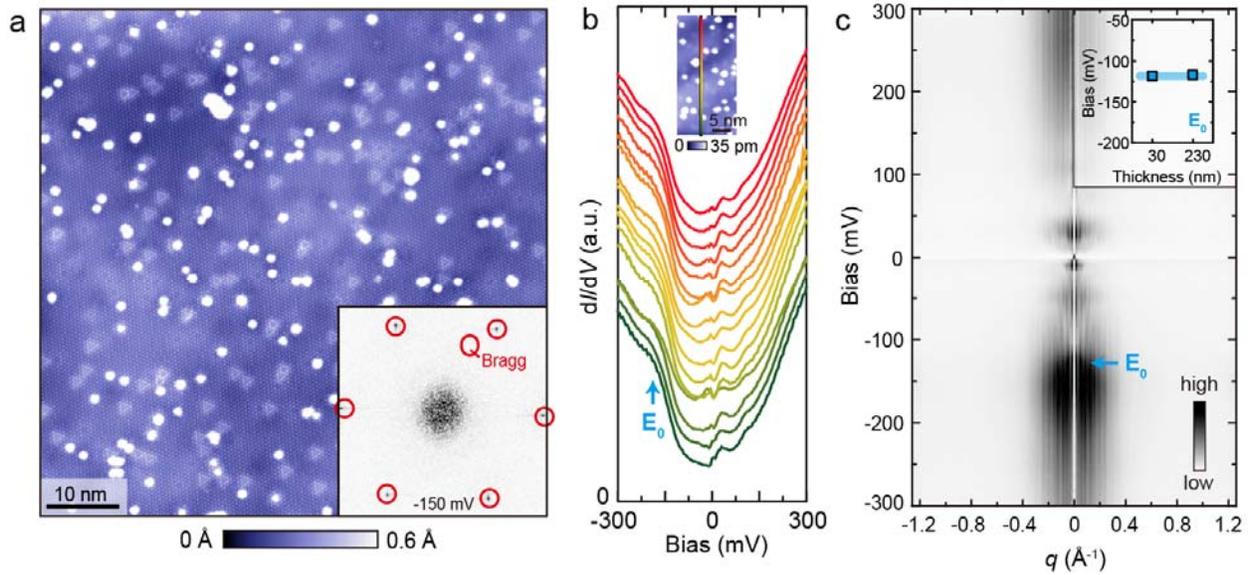

**Figure 2. Large-scale electronic structure. a** Typical STM topograph of a large flat region of the $Fe_3Sn$ surface. Inset in a shows Fourier transform of a normalized d$I$/d$V$(**r**, $V$) map ((d$I$/d$V$)/($I$/$V$)) for bias V = -150 mV. Atomic Bragg peaks related to the *ab*-plane lattice vectors are circled in red. **b** d$I$/d$V$ linecut acquired over the vertical line denoted in the STM topograph in the inset. Blue arrow labeled as $E_0$ denotes an abrupt increase in conductance identified as a partial flat band in Ref. [28]. **c** Radially averaged linecut generated from the FTs of normalized d$I$/d$V$(**r**, $V$) maps ((d$I$/d$V$)/($I$/$V$)). $E_0$ denotes the energy below which the magnitude of the QPI wave vector is notably enhanced, concomitant with the increase in conductance in **b**. Inset in **c** shows experimentally determined $E_0$ for two film thicknesses extracted from the equivalent radially averaged linecuts as shown in **c**. STM setup conditions: **a** $I_{set}$ = 96 pA, $V_{sample}$ = 12 mV; **inset in a, b-c** $I_{set}$ = 1 nA, $V_{sample}$ = 300 mV, $V_{exc}$ = 5 mV.

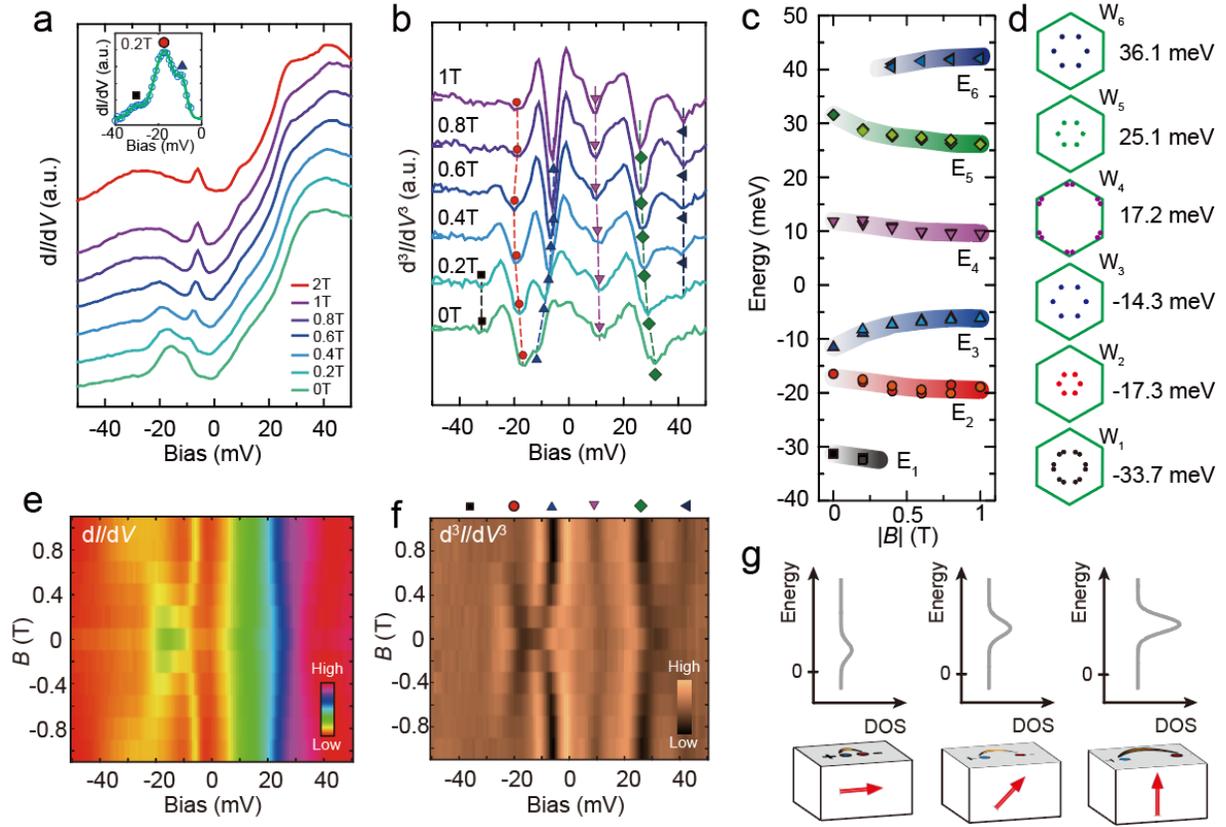

**Figure 3. Magnetization-direction-driven tunability of the prominent spectral features. a** Average d*I*/d*V* spectra and **b** second derivative of the spectra (d³*I*/d*V*³) acquired in a magnetic field applied parallel to the *c*-axis, offset for clarity. Symbols in (b) denote local minima in d³*I*/d*V*³ curves, corresponding to the peaks/kinks in the equivalent d*I*/d*V* spectra. Dashed lines serve as guides to the eye for the dispersions of local minima in field. Short bars on the left-bottom of each curve in (b) indicate the zero value for each curve. **c** Dispersion of spectral peaks in magnetic field. The symbols in darker (brighter) colors are extracted from the data taken in positive (negative) fields. **d** Theoretically calculated Weyl node positions for magnetization saturated along the c-axis [27]. Waterfall plots of **e** d*I*/d*V* and **f** d³*I*/d*V*³ as a function of magnetic field applied parallel to the *c*-axis, including the data acquired in negative fields (antiparallel to the *c*-axis), presented as intensity maps in different color scales for a better visualization. **g** Schematic of one scenario of Weyl node evolution and its effect on the density of states.

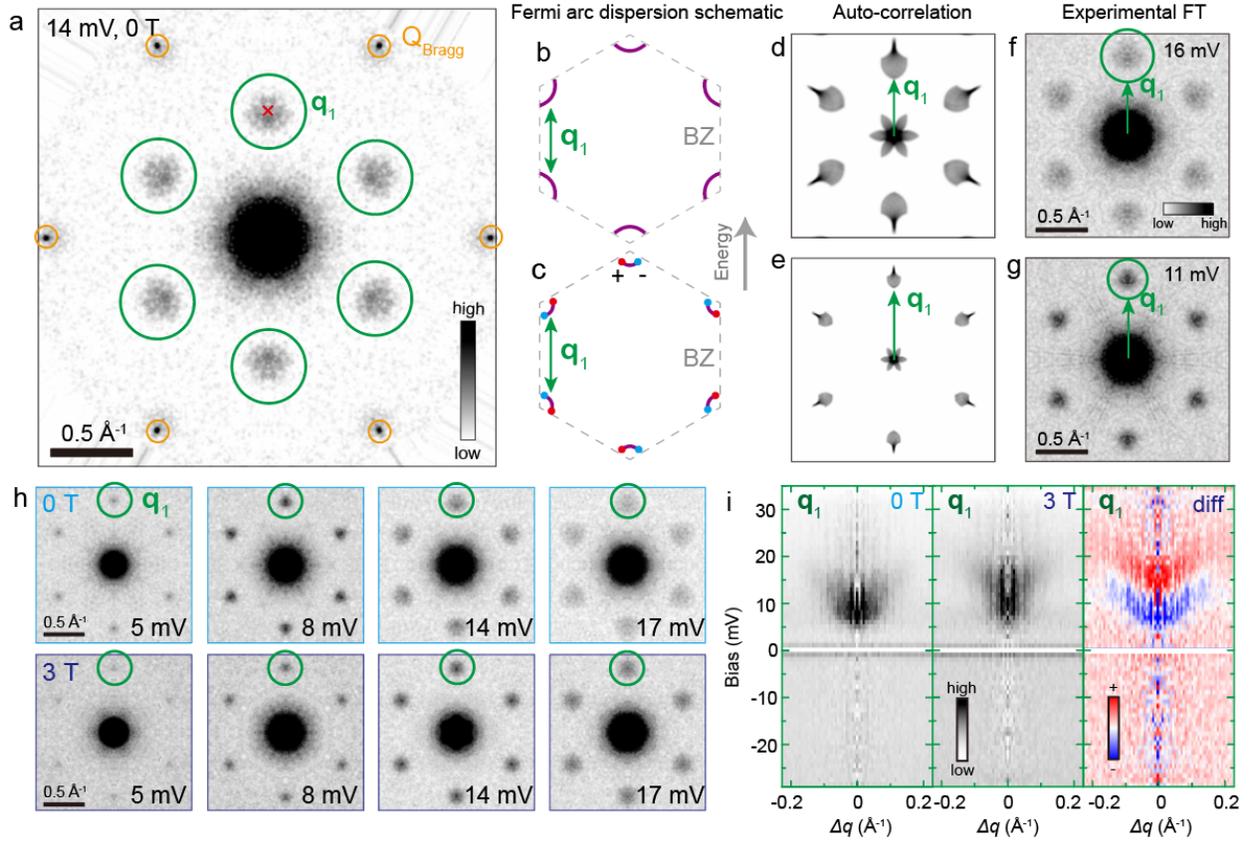

**Figure 4. Field-tunable signature of electron scattering and interference. a** Six-fold symmetrized FT of L(**r**,V) = (dI/dV)/(I/V) map at 14 mV. Orange circles enclose lattice Bragg peaks **Q**$_{Bragg}$. The main scattering wave vectors are circled in green (**q**$_1$). Constant energy contour schematic of Fermi arc dispersion with **b** large and **c** small Fermi arcs originating from pairs of Weyl nodes around each K point (red and blue circles in **c**), and **d-e** corresponding auto-correlations. **f,g** Experimental L(**r**,V) maps at 16 mV and 11 mV, respectively, showing the scattering signature qualitatively consistent with those in **d,e**. **h** Comparison of L(**r**,V) maps at zero field (top row) and 3 T (bottom row). **i** Radially averaged linecuts of FTs of L(**r**,V) maps, starting at (**Q**$_{Bragg,1}$+**Q**$_{Bragg,2}$)/3, focusing on the reciprocal space region denoted by the green circle in **a**, at 0 T (left), 3 T (middle) and their difference (right). STM setup conditions: $I_{set}$ = 280 pA, $V_{sample}$ = 35 mV, $V_{exc}$ = 1 mV.

14